\documentclass[reprint,
amsmath,amssymb,
aps,
]{revtex4-1}

\makeatletter
\let\frontmatter@footnote@produce\frontmatter@footnote@produce@endnote
\makeatother

\usepackage{graphicx}
\usepackage{dcolumn}
\usepackage{bm}
\usepackage{amssymb,physics}
\usepackage{braket}
\usepackage[colorlinks,allcolors=blue]{hyperref}

\def\br{\bm{r}}
\def\bk{\bm{k}}
\def\bkp{\bm{k}'}
\def\bq{\bm{q}}

\def\bG{\bm{G}}

\begin{document}

\title{Efficient \textit{Ab Initio} Calculations of Electron-Defect Scattering\\ and Defect-Limited Carrier Mobility}

\author{I-Te Lu}
\author{Jin-Jian Zhou}%
\author{Marco Bernardi}
\email{bmarco@caltech.edu}
\affiliation{Department of Applied Physics and Materials Science, California Institute of Technology, Pasadena, California 91125}

\date{\today}

\begin{abstract}
Electron-defect ($e$-d) interactions govern charge carrier dynamics at low temperature, where they limit the carrier mobility and give rise to phenomena of broad relevance in condensed matter physics.
\textit{Ab initio} calculations of $e$-d interactions are still in their infancy, mainly because they require large supercells and computationally expensive workflows. 
Here we develop an efficient \textit{ab initio} approach for computing elastic $e$-d interactions, their associated $e$-d relaxation times (RTs), and the low-temperature defect-limited carrier mobility. 
The method is applied to silicon with simple neutral defects, such as vacancies and interstitials. 
Contrary to conventional wisdom, the computed $e$-d RTs depend strongly on carrier energy and defect type, and the defect-limited mobility is temperature dependent. 
These results highlight the shortcomings of widely employed heuristic models of $e$-d interactions in materials. 
Our method opens new avenues for studying $e$-d scattering and low-temperature charge transport from first principles. 
\end{abstract}

\maketitle

\section{Introduction}
%
%
Defects in materials can scatter or capture charge carriers. They control carrier dynamics at low temperature, where phonons are frozen out, or even at room temperature in the presence of strong disorder.
Impurity defects are key to controlling the carrier mobility in semiconductors~\cite{Radisavljevic2013,Wang2012,Li2016,Herz2017,Mao2017,Sun2015}, 
and defects are widely employed to engineer advanced materials and devices~\cite{Xu2017,Koenraad2011,Gunlycke2011,Yazyev2010}. 
Weak localization \cite{Bergmann, WAL-Lee, Datta} and conductance fluctuations~\cite{UCF-Lee, akkermans2007mesoscopic} are but two examples of subtle quantum effects that can be induced by defects. 
For all these reasons, understanding how defects interact with charge carriers is important in both applied and fundamental condensed matter physics.\\
\indent
%
%
Calculations of electron-defect ($e$-d) interactions have so far relied almost exclusively on empirical models~\cite{Chattopadhyay1981}, in which the atomic details of the defects are suppressed and simple approximations are employed for the electronic screening and band structure.
For example, the perturbation induced by charged defects is typically modeled with a simple Yukawa potential, and models of $e$-d scattering due to neutral (non-charged) defects are scarce and mainly limited to spherical-well-potential approaches~\cite{Erginsoy1950,Sclar1956,Blakemore1980}.
Yet, the perturbation induced by defects in real materials can be anisotropic and complex, and its effects on electronic screening intricate. 
How the band structure and defect states impact $e$-d scattering is difficult to predict.\\
\indent
Due to their severe approximations, empirical models are not reliable to compute $e$-d interactions, and their predictions can be qualitatively and quantitatively incorrect. 
First-principles approaches, such as those based on density functional theory (DFT) and related methods~\cite{martin2004electronic}, provide quantitative accuracy in many areas of materials physics, and are therefore desirable to treat $e$-d interactions.
%
%
%
Such \textit{ab initio} $e$-d calculations can take advantage of existing tools developed for electron-phonon ($e$-ph) interactions, for which accurate calculations of relaxation times (RTs)~\cite{Bernardi2014,Bernardi2015a,Bernardi2015b,Tanimura2016}, matrix elements and their interpolation~\cite{Agapito2018}, and phonon-limited carrier mobility~\cite{Mustafa2016, Park2014, Zhou2016, Lee2018, Zhou2018} have been recently developed. These and other workflows developed for $e$-ph calculations can be generalized to treat $e$-d interactions.\\ 
\indent
%
%
However, there are open challenges specific to $e$-d calculations that currently prevent their broad applicability. 
First-principles $e$-d calculations need large supercells to obtain the electron wavefunctions and defect perturbation potentials; 
they additionally involve computationally costly $e$-d matrix elements, and require systematic convergence of the RTs with respect to supercell size and Brillouin zone (BZ) grids. 
Early work on first-principles $e$-d interactions~\cite{Evans2005,Restrepo2009,Lordi2010} proposed an approach, called here the \textit{all-supercell method}, 
in which the cost of computing $e$-d matrix element is prohibitively large due to the use of electron wavefunctions from large supercells. 
This work aims to address these open challenges and make \textit{ab initio} $e$-d calculations more affordable and broadly applicable.\\ 
\indent
%
%
Here we formulate an efficient scheme to compute and converge the $e$-d matrix elements and the associated RTs and defect-limited carrier mobility. 
Our approach does not require the wavefunctions of large supercells, which dramatically speeds up the calculations. 
We apply our method to study $e$-d interactions in silicon in two cases, neutral vacancy and interstitial defects, 
for which we compute and converge the $e$-d RTs as a function of energy and the carrier mobility as a function of temperature below 150 K.
The results show that, contrary to conventional wisdom, the RTs depend strongly on carrier energy and defect type, and the defect-limited mobility for neutral defects depends on temperature.
Our results provide new microscopic insight into $e$-d scattering, and our approach, together with its future extensions, can uncover new defect physics in materials and devices for electronics, energy and quantum technologies.\\ 
\indent 
The paper is organized as follows:
Section~\ref{sec:methodology} describes the new formulas and workflow for computing $e$-d matrix elements and RTs, and compares our approach with the all-supercell method. 
In Section~\ref{sec:results} we compute and converge $e$-d RTs for electrons and holes in silicon and the associated defect-limited mobility at low temperature.  
Section~\ref{sec:discussion} discusses technical points and outlines future research directions. 
%

\section{Methodology}
\label{sec:methodology}
%
%
In the following, we work under the assumption that the defects are neutral (non-charged) and that the $e$-d scattering events are independent, uncorrelated and elastic. 
The $e$-d scattering rate $\Gamma_{n\bk}$ (and its inverse, the RT, $\tau_{n\bk} = \Gamma_{n\bk}^{-1}$) for a Bloch state $\ket{n\bk}$, where $n$ is the band index and $\bk$ the crystal momentum, 
is computed using lowest-order perturbation theory (see Appendix~\ref{sec:RTderivation}): 
\begin{equation}\label{eq:incoherent}
\Gamma_{n\bk}=\frac{2\pi}{\hbar}\frac{n_{\text{at}}C_{\rm d}}{N_{\bkp}}\sum_{n'\bkp}\left|M_{n'\bkp,n\bk}\right|^{2}\delta\left(\varepsilon_{n'\bkp}-\varepsilon_{n\bk}\right),
\end{equation}
where $n_{\text{at}}$ is the number of atoms in a primitive cell, $C_{\rm d}$ the (dimensionless) defect atomic concentration,  
$N_{\bkp}$ the number of BZ $\bkp$-points used in the sum, 
and $\varepsilon_{n\bk}$ the unperturbed energy of the state $\ket{n\bk}$. 
The $e$-d matrix elements $M_{n'\bkp,n\bk}$ are the central quantities computed in this work; they encode the probability amplitude for scattering from the unperturbed state $\ket{n\bk}$ to $\ket{n'\bkp}$ 
due to the perturbation potential $\Delta V_{\rm e-d}$ from a defect:
\begin{equation}\label{eq:matrixelements}
M_{n'\bkp,n\bk}=\braket{ n' \bkp \,|\, \Delta V_{\rm e-d}\, |\, n\bk }.
\end{equation}
\indent
Within DFT~\cite{Kohn1965}, the $e$-d perturbation can be computed as the difference between the Kohn-Sham (KS) potentials 
$V_{\rm KS}$ of a defect-containing supercell and a pristine supercell with no defect, namely, $\Delta V_{\rm e-d} = V_{\rm KS}^{\rm (d)} - V_{\rm KS}^{\rm (p)}$. 
Here and below, we use superscripts (d) and (p) to denote the defect-containing and pristine systems, respectively. 
%
%
When using, as we do here, norm-conserving pseudopotentials in the Kleinman-Bylander (KB) form~\cite{Kleinman1982}, the Kohn-Sham potential can be written as a sum of local and nonlocal parts~\cite{martin2004electronic}:
\begin{equation} 
V_{\rm KS} = V_{\rm L}(\br) + \hat{V}_{\rm NL}.
\label{eq:KS}
\end{equation}
The local potential $V_{\rm L}(\br)$ comprises the Hartree and exchange-correlation potentials plus the local part of the pseudopotentials,
\begin{equation} 
V_{\rm L}(\br) = V_{\rm H}(\br) + V_{\rm XC}(\br) + V_{\rm pp}(\br). 
\end{equation}
The nonlocal potential $\hat{V}_{\text{NL}}$, which is due to the pseudopotentials, is as a sum over all atoms in the supercell of KB projectors $\ket{\beta_{i}^{(s)}}$, each localized in the core region of atom $s$: 
\begin{equation}\label{eq:vnl}
\hat{V}_{\text{NL}} = \sum_{s=1}\sum_{ij}D_{ij}^{(s)}\ket{\beta_{i}^{(s)}}\bra{\beta_{j}^{(s)}}, 
\end{equation}
where $i$ and $j$ are orbital angular momentum quantum numbers, and $D_{ij}^{(s)}$ are KB coefficients~\cite{martin2004electronic}.\\
\indent
%
%
Accordingly, we separate the matrix elements into a local and a nonlocal part, 
\begin{equation}\label{eq:2mels}
M_{n'\bkp,n\bk}=M_{n'\bkp,n\bk}^{\text{L}} + M_{n'\bkp,n\bk}^{\text{NL}}\,\,,
\end{equation}
each due to the respective defect perturbation, $\Delta V_{\rm L} (\br)= V_{L}^{\rm (d)} (\br)- V_{L}^{\rm (p)}(\br)$ for the local and $\Delta \hat{V}_{\rm NL} = \hat{V}_{\rm NL}^{\rm (d)} - \hat{V}_{\rm NL}^{\rm (p)}$ for the nonlocal part:
\begin{equation}
\begin{split}
M_{n'\bkp,n\bk}^{\text{L}} &= \braket{ n' \bkp \,|\, \Delta V_{\rm L}(\br) \,|\, n \bk } \\ 
M_{n'\bkp,n\bk}^{\text{NL}} &= \braket{ n' \bkp \,|\, \Delta \hat{V}_{\rm NL} \,|\, n \bk }.
\end{split}
\end{equation}
%
\begin{figure*}[ht]
\centering 
\includegraphics[width=0.9\linewidth]{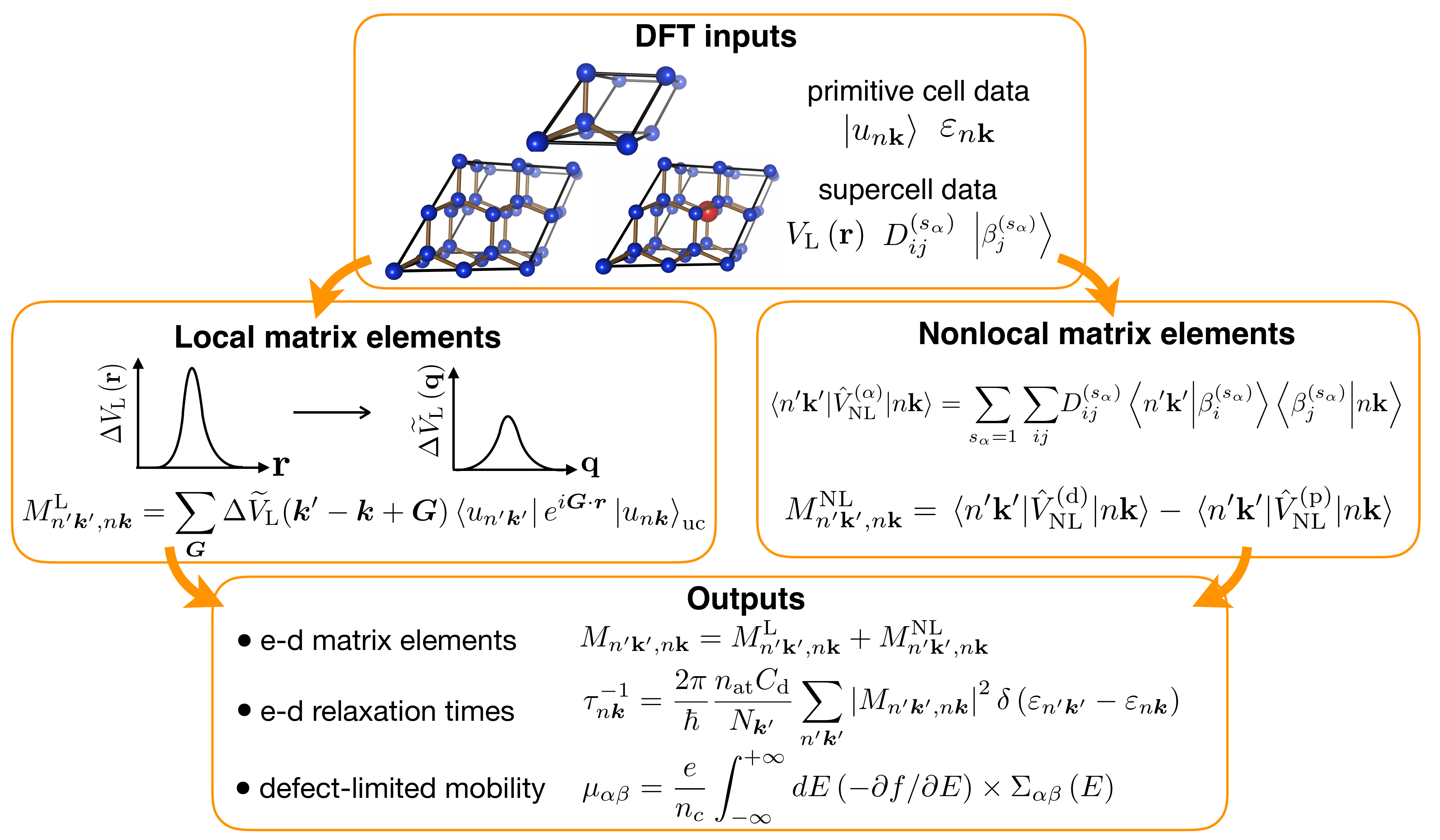}
\caption{ Workflow for computing the $e$-d matrix elements and relaxation times, and the defect-limited mobility. In the first step, several inputs are computed with DFT, including the KS wavefunctions and eigenvalues of a primitive cell, 
and the local and nonlocal parts of the KS potential, separately in a pristine and a defect-containing supercell. 
The local matrix elements are then computed by Fourier transforming and interpolating the local perturbation potential, $\Delta V_{\rm L}(\br)$, and combining it with the plane-wave matrix elements of the primitive cell. 
The nonlocal matrix elements are similarly computed by splitting the calculation into a primitive cell and a supercell part, using only the KS wavefunctions of the primitive cell. 
The total $e$-d matrix elements are then formed by adding the local and nonlocal parts, following which they are employed to compute as output the properties of interest, 
including the band and momentum-resolved $e$-d relaxation times and the defect-limited mobility. 
}\label{fig:workflow}
\end{figure*}
%
%
\subsection{Electron-defect matrix element computation}
\vspace{-10pt}
We develop a new approach to reduce the computational cost of the local and nonlocal matrix elements. 
We Fourier transform the local perturbation potential $\Delta V_{\text{L}}(\br)$ and compute the local matrix elements as (see Appendix~\ref{sec:LocMderivation})
\begin{equation}\label{eq:localM}
M_{n'\bkp,n\bk}^{\text{L}}=\sum_{\bG}\Delta \widetilde{V}_{\rm L} (\bkp-\bk +\bG) \braket{ u_{n'\bkp} |\, e^{i\bG\cdot\br} \,| {u_{n\bk}} }_{\rm uc},
\end{equation}
where $u_{n\bk}(\br)$ is the periodic part of the Bloch wavefunction (normalized in a primitive cell with volume $\Omega_{\text{uc}}$), 
$\bG$ are reciprocal lattice vectors of the primitive cell, and $\Delta \widetilde{V}_{\rm L}$ are Fourier coefficients of the local perturbation potential (computed in a supercell with volume $\Omega_{\text{sup}}$): 
\begin{equation}\label{eq:VlocFT}
\Delta \widetilde{V}_{\rm L} (\bq)=\frac{1}{\Omega_{\rm uc}} \int_{\Omega_{\text{sup}}}\!\! d\br\,\, \Delta V_{\rm L}\left(\br\right)e^{-i\bq\cdot\br},
\end{equation}
where $\bq = \bkp - \bk + \bG$ is the transferred momentum in the $\ket{n\bk} \rightarrow \ket{n'\bkp}$ scattering process. 

Through Eq.~(\ref{eq:localM}), we effectively separate the local matrix element calculation into two independent parts, 
the plane-wave matrix elements $\braket{ u_{n'\bkp} |\, e^{i\bG\cdot \br} \,| {u_{n\bk}} }_{\rm uc}$,  
which are easily computed by integrating over the \textit{primitive} cell, and the Fourier coefficients $\Delta \widetilde{V}_{\rm L} (\bq)$, which are computed in the supercell. 
In addition, since the local perturbation potential is smooth and decays in real space over a few angstroms, its Fourier coefficients $\Delta \widetilde{V}_{\rm L}(\bq)$ decay rapidly in reciprocal space, 
and the summation over $\bG$ in Eq.~(\ref{eq:localM}) can be truncated to a small cutoff $-$ typically, just a few reciprocal lattice vectors. 
For the same reason, the Fourier coefficients $\Delta \widetilde{V}_{\rm L}(\bq)$ can be efficiently interpolated 
(in our case, with a B-spline) at arbitrary $\bq$ starting from their calculation at a few thousand $\bq$-points in a cubic box centered at $\bq=0$. 
A great advantage of this formulation is that one can compute the local matrix elements using only the wavefunctions of the primitive cell (as opposed to those of the supercell), 
and effectively interpolate the perturbation potential to fine BZ grids.\\ 
\indent
%
%
The nonlocal matrix elements $M_{n'\bkp,n\bk}^{\text{NL}}$ are computed as the difference between the nonlocal potentials of a defect-containing and a pristine supercell: 
\begin{equation}\label{eq:totnonlocal}
\begin{split}
M_{n'\bkp,n\bk}^{\rm NL} &= \braket{n'\bkp | \hat{V}_{\rm NL}^{\rm (d)} - \hat{V}_{\rm NL}^{\rm (p)} | n\bk} \\ 
&= \braket{n'\bkp | \hat{V}_{\rm NL}^{\rm (d)} | n\bk}  -   \braket{n'\bkp | \hat{V}_{\rm NL}^{\rm (p)} | n\bk}, 
\end{split}
\end{equation}
where the matrix elements of $\hat{V}_{\rm NL}$ for each of the two supercells (labelled by $\alpha =$ d, p), using Eq.~(\ref{eq:vnl}), read: 
\begin{equation}\label{eq:nonlocalM}
\begin{split}
\!\!\!\braket{n'\bkp | \hat{V}_{\rm NL}^{\rm (\alpha)} | n\bk} = \sum_{s_\alpha=1}\sum_{ij}&D_{ij}^{(s_\alpha)} \braket{n'\bkp | \beta_{i}^{(s_\alpha)}} \braket{ \beta_j^{(s_\alpha)} | n\bk}.
\end{split}
\end{equation}
Similar to the local matrix elements, the computation in Eq.~(\ref{eq:nonlocalM}) is split into a primitive cell and a supercell calculation, by expressing the scalar products $\braket{\beta_j^{(s)} | n\bk }$ as (see Appendix~\ref{sec:NonLocMderivation})
\begin{equation}\label{eq:scalarproduct} 
\braket{\beta_j^{(s)} | n\bk } = \frac{1}{\sqrt{\Omega_{\rm uc}}} \sum_{\bG} B_{j\bk}^{(s)*}( \bG ) \braket{ e^{ i\bG \cdot \br } | u_{n\bk} }_{\rm uc},
\end{equation}
where $B_{j\bk}^{(s)}\left(\bG\right)$ is the Fourier coefficient of the KB projector $\beta_{j}^{(s)}$ (multiplied by the phase factor $e^{-i\bk \cdot \br}$) 
at the primitive-cell reciprocal lattice vector $\bG$: 
\begin{equation}\label{eq:NLFT}
B_{j\bk}^{(s)} (\bG) = \frac{1}{\sqrt{\Omega_{\rm uc}}} \int_{\Omega_{\rm sup}}\!\!d\br\,\, \left(\beta_j^{(s)}(\br) e^{-i\bk\cdot\br} \right)e^{-i\bG\cdot\br}.
\end{equation}
Note that the nonlocal matrix elements, which are computed as the difference in Eq.~(\ref{eq:totnonlocal}), 
are nonzero both because the atomic positions change upon relaxing the structure in the defect-containing supercell 
and because the number and type of atoms, in general, differ in the two supercells, as is the case when considering a vacancy or impurity.\\
\indent
%
%
Figure~\ref{fig:workflow} shows our workflow for computing properties related to $e$-d interactions from first principles.  
To obtain the relevant DFT inputs, the KS equations are solved in a primitive cell and separately in pristine and defect-containing supercells. 
The local and nonlocal matrix elements are then computed by splitting the calculation into a primitive cell and a supercell part, an approach that dramatically reduces the computational cost. 
Computing local matrix elements is the most expensive step, while the nonlocal matrix elements only involve relatively inexpensive reciprocal space sums.  
Importantly, only the KS wavefunctions, band structure and $\bk$-points of the primitive cell are used, while the supercells are employed only to obtain the perturbation potential due to the defect. 
Once computed, the $e$-d matrix elements are employed to calculate the $e$-d RTs and the defect-limited mobility, among other quantities of interest. 
This approach allows us to systematically converge the RTs and other properties related to $e$-d interactions with respect to supercell size and BZ grids.

\subsection{Comparison with the all-supercell method}
\begin{figure}[!hb]
\centering
\includegraphics[width=0.95\linewidth]{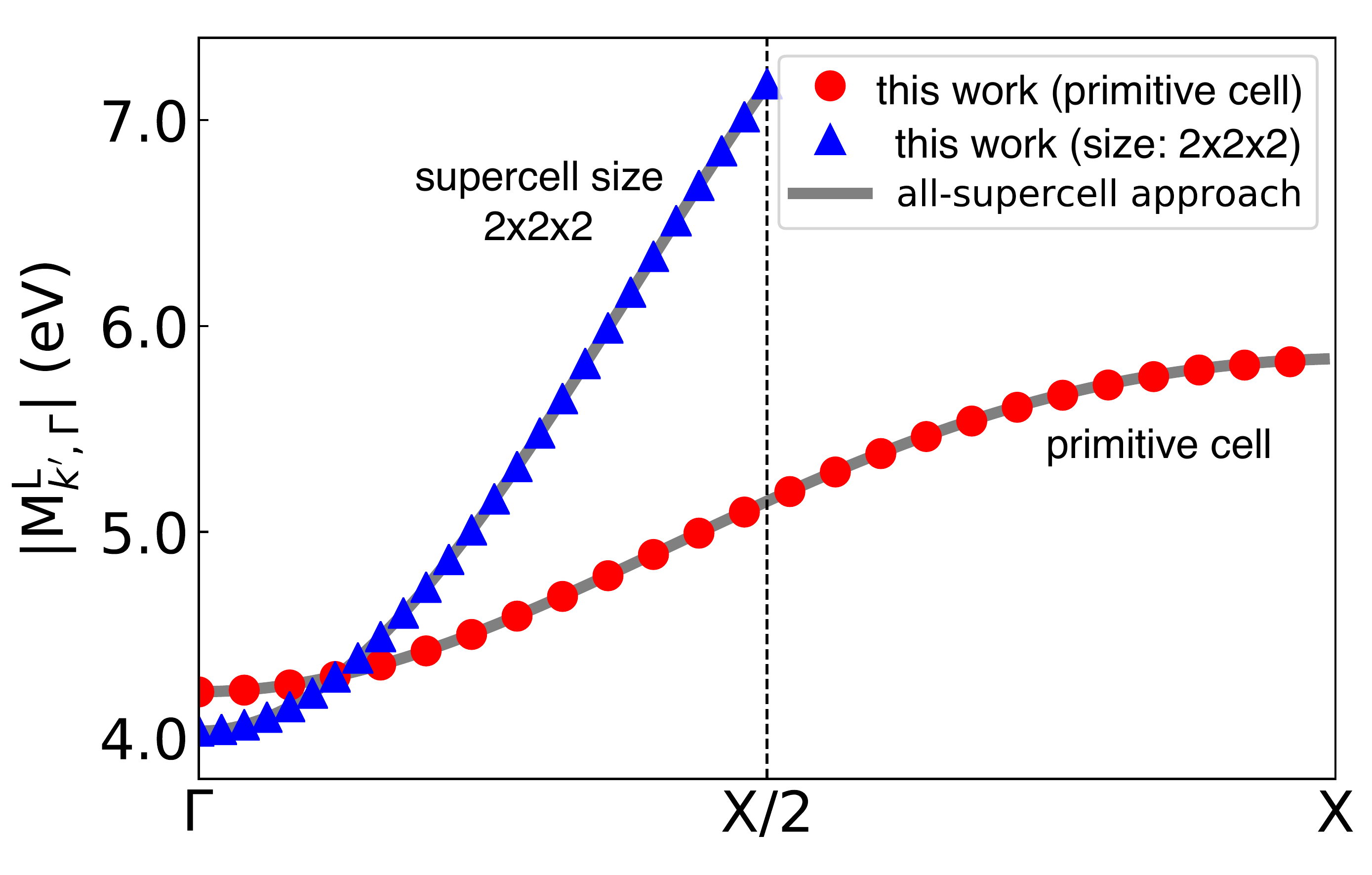}
\caption{
Absolute value of the local $e$-d matrix elements, obtained from our approach and, for comparison, with the all-supercell method, for two cell sizes, a primitive cell and a $2~\times~2~\times~2$ supercell. 
The initial state is in the lowest valence band at $\Gamma$, and the final states are in the same band with crystal momenta $\bkp$ along the $\Gamma$$-$$X$ high-symmetry line.
}\label{fig:Methods_comparison}
\end{figure}

\vspace{-10pt}
In the all-supercell method~\cite{Restrepo2009}, one uses the pristine and defect-containing supercells to provide all the necessary quantities for computing the $e$-d matrix elements, 
including the wavefunctions, band structure, perturbation potentials, and BZ grids. 
However, using supercell wavefunctions makes it challenging to compute and converge the $e$-d matrix elements and RTs, 
and ultimately to carry out accurate $e$-d calculations, since unconverged $e$-d RTs and transport properties can differ widely from the converged results.\\
\indent
All $e$-d calculations need to be converged with respect to supercell size; as we discuss below, converging the RTs for a neutral defect typically requires very large supercells with hundreds of atoms.  
In our approach, this convergence does not constitute a challenge since the same (primitive cell) wavefunctions are employed, regardless of supercell size. 
Conversely, in the all-supercell method, one uses wavefunctions from the pristine supercell, 
and the computational cost to compute and store the wavefunctions and obtain the matrix elements increases dramatically with supercell size, 
making accurate convergence tests too computationally demanding.\\ 
\indent 
Let us analyze the cost of the most computationally intensive step, namely obtaining the local $e$-d matrix elements, $M^{\rm L}_{n'\bkp,n\bk}$. 
Using a uniform BZ grid with $N_{\bk}$ points, one obtains $\mathcal{O}(N_{\bk}^2)$ matrix elements, 
each for a distinct $\ket{n\bk} \rightarrow \ket{n'\bkp}$ $e$-d scattering process. 
In a typical calculation, a uniform grid with at least $N_{\bk} \approx 10^6$ points is needed to converge the RTs in the entire BZ.  
In a mobility calculation, one typically selects a small energy window of $\sim$100 meV near the band edges (in a semiconductor, or near the Fermi energy in a metal), 
which reduces the required number of $\bk$-points to $N_{\bk}\approx10^4$.\\
\indent
In the all-supercell method, the local matrix elements are computed as:
\begin{equation}\label{eq:localMsup}
M^{\rm L}_{n'\bkp,n\bk} = \braket{n'\bkp | \Delta V_{\rm L}(\br) | n\bk}_{\rm sup}
\end{equation}
where the subscript (sup) denotes that both the local perturbation potential $\Delta V_{\rm L}(\br)$ and the wavefunctions are obtained from a DFT calculation on a supercell. 
Since the cost of the DFT calculations scales with system size as roughly $N^{3}_{\rm at}$, where $N_{\rm at}$ is the number of atoms in the supercell, computing $N_{\bk}$ supercell wavefunctions, from which the local matrix elements are computed on the uniform grid, costs $N_{\bk} N^{3}_{\rm at}$ in the all-supercell method.\\ 
\indent
By contrast, in our method only the primitive cell wavefunctions are used, and thus the computational cost of the matrix elements does not depend on $N_{\rm at}$ through the wavefunctions. 
To obtain the local matrix elements on the uniform grid with our method [see Eq.~(\ref{eq:localM})], the only supercell data one needs are the Fourier coefficients $\Delta \widetilde{V}_{\rm L} (\bq)$ of the local perturbation potential. 
Obtaining these coefficients at a few thousand $\bq$ points $-$ from which an interpolation table can be constructed $-$ has a cost that scales as $N^{3}_{\rm at}$, 
but this step is required only once for a given supercell size. Therefore, computing the local matrix elements on a uniform BZ grid with $N_{\bk}$ points has a cost of order $N^{3}_{\rm at}$ in our method, 
versus a cost of $N_{\bk}N^{3}_{\rm at}$ in the all-supercell method. 
For the typical mobility calculation mentioned above, this represents a speed-up by a factor of $N_{\bk}\approx 10,000$ over the all-supercell method. 
Note that our mobility calculations are already expensive (tens of thousands of core-hours), so approaches that are thousands of times more expensive are clearly impractical. 
An additional benefit is that in our approach the large supercell wavefunctions are never stored or loaded into memory, so the speed up is significant even for computing a single $e$-d RT.\\
\indent
Finally, one would like to map the $e$-d scattering processes onto the band structure of the primitive cell, as is done for $e$-ph scattering processes. 
This is possible in our approach due to our use of primitive cell band structures and $\bk$-point grids, but impractical in the all-supercell method, where one uses the supercell band structures and $\bk$-point grids, 
which depend on the choice of a supercell and differ from those of the primitive cell due to nontrivial BZ folding effects. 
Due to its computational efficiency and convenience, we thus believe that our approach solves key technical challenges that have so far prevented efficient and accurate \textit{ab initio} calculations of $e$-d interactions.\\
\indent
Figure~\ref{fig:Methods_comparison} validates our approach, by comparing the local matrix elements computed with our method [using Eq.~(\ref{eq:localM})] 
with those obtained with the all-supercell method using Eq.~(\ref{eq:localMsup}), for vacancy defects in silicon 
(see below for the computational details). It is seen that for a test case of a primitive cell and a 2$\,\times\,$2$\,\times\,$2 supercell the results obtained with the two methods are in perfect agreement. This is but one of many benchmark tests we have performed. 
%
%
\newpage
\section{Results}
\label{sec:results}
\subsection{Computational details}
\vspace{-10 pt}
We apply our approach to compute the $e$-d RTs and defect-limited mobility in silicon, separately for vacancy and (tetrahedral) interstitial defects.  
A defect concentration of 1 ppm (one defect in 10$^6$ atoms) is assumed in both cases. 
The ground states of the primitive cell and of supercells with size $N\,\times\,N \,\times\, N$ (where $N$ is the number of primitive cells along each lattice vector) are computed using DFT within the local density approximation~\cite{Perdew1981}, using a plane wave basis and norm-conserving pseudopotentials~\cite{Kleinman1982} with the {\sc Quantum Espresso} package~\cite{Giannozzi2009}. 
Briefly, for the primitive cell we use a lattice constant of $5.43$ \r{A}, a $40$ Ry kinetic energy cutoff and a $12\,\times\,12 \,\times\, 12$ Monkhorst-Pack $\bk$-point grid~\cite{Monkhorst1976}, 
converging the total energy to within $10$ meV/atom; a consistent lattice constant and total energy convergence criterion is employed for the supercells.
%
In the defect-containing supercells, the atomic forces are relaxed to within 25 meV/{\AA} to account for the structural changes induced by the defect, and the resulting KS potentials are used to compute the $e$-d matrix elements. 
Due to the different reference potentials in the pristine and defect-containing supercells,
we employ the core-average potential alignment method~\cite{Kumagai2014} to align the local potentials of the two supercells when computing the local perturbation potential; 
the reference potential is taken as the average of the local potential at the atom that is farthest from the defect site.
In the $e$-d RT and mobility calculations, we select the electronic states of relevance in a small ($\sim$100 meV) energy window near the band edges, 
and interpolate the band structure using maximally localized Wannier functions~\cite{Marzari1997} with the {\sc Wannier90} code~\cite{Yates2007,Mostofi2014}. 
All $e$-d calculations have been implemented in our {\sc perturbo} code~\cite{Perturbo} following the workflow in Fig.~\ref{fig:workflow}. 
%

\begin{figure}[!t]
\centering
\includegraphics[width=0.9\linewidth]{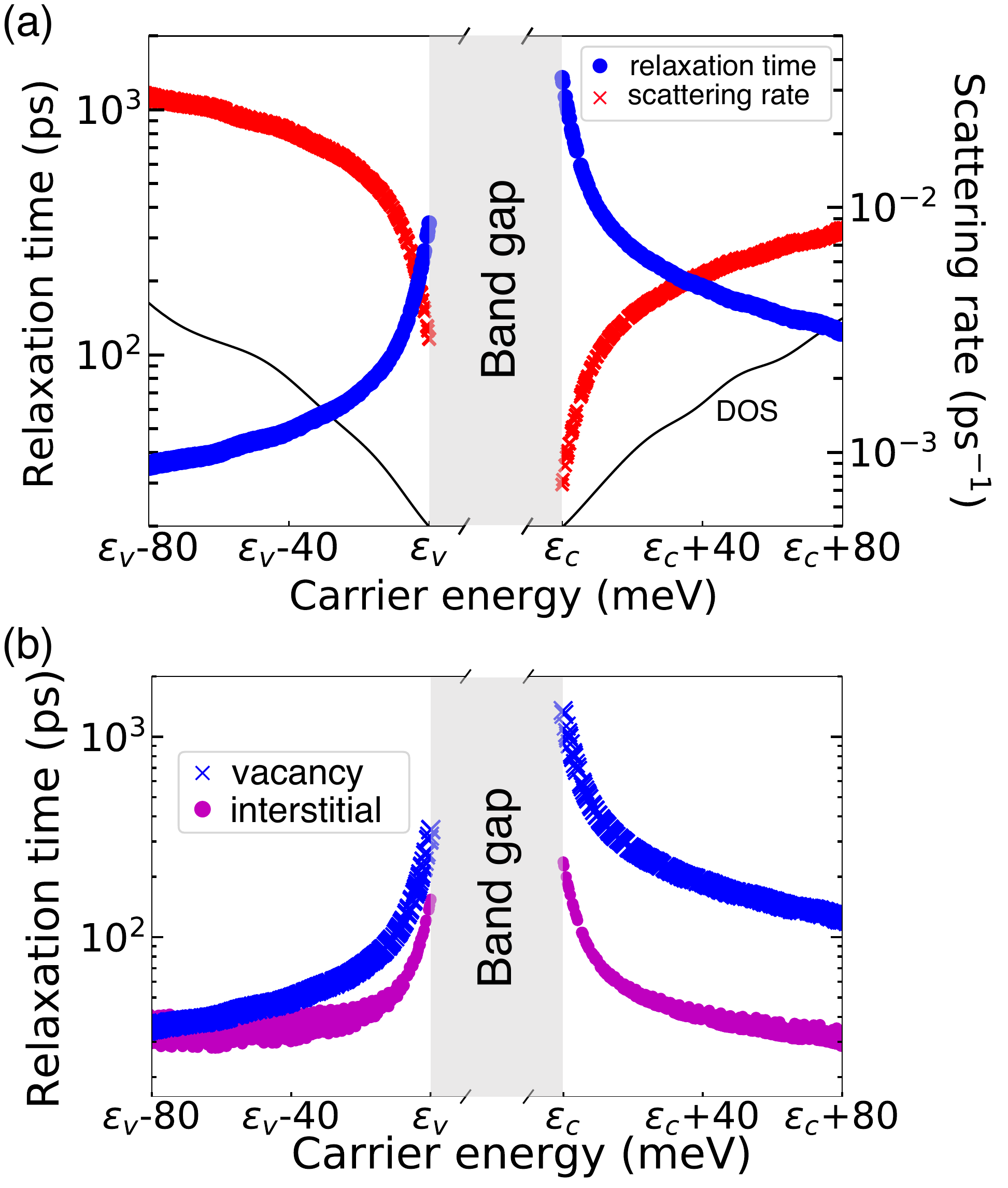}
\caption{(a) The relaxation times and their inverse, the scattering rates, for electrons and holes due to $e$-d scattering with vacancy defects in silicon. 
The valence band maximum is labeled $\varepsilon_{v}$ and the conduction band minimum $\varepsilon_{c}$. 
The density of states (DOS) is also plotted, in arbitrary units.
(b) Comparison between the relaxation times due to $e$-d scattering with vacancy and interstitial defects in silicon. 
The calculations use a $200^{3}$ BZ grid with a $5$ meV broadening, and a supercell size of $6\times6\times6$ (432 atoms) for vacancy and $8\times8\times8$ (1024 atoms) for interstitial defects.}
\label{fig:first_time_edRT}
\end{figure}

\begin{figure*}[t!]
\centering
\includegraphics[width=1.03\linewidth]{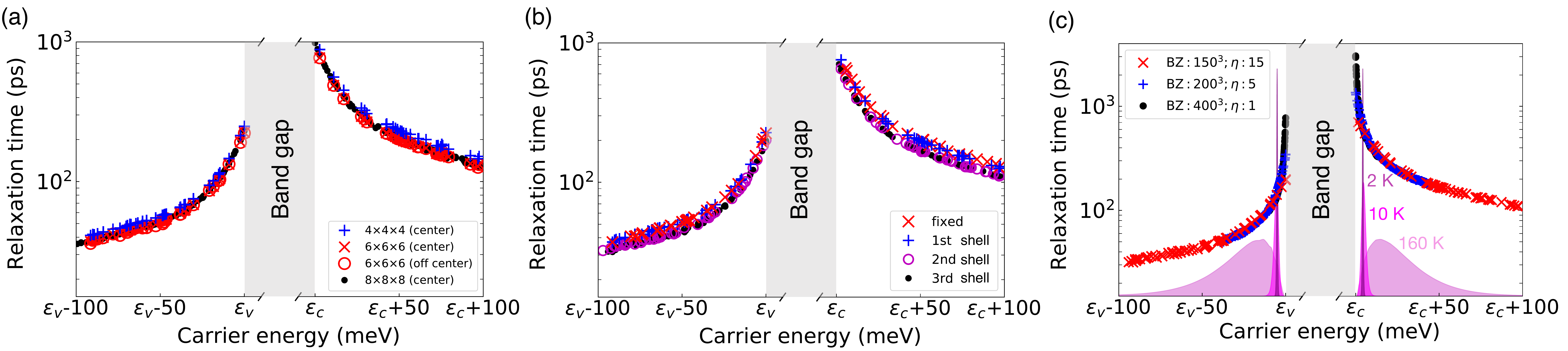}
\caption{
Convergence of the $e$-d RTs, shown here for vacancy defects in silicon.
We consider the effect of (a) supercell size,  
(b) structural relaxation, and (c) BZ grid and energy broadening in Eq.~(\ref{eq:incoherent}), where for each broadening value we use 
a converged BZ grid. 
Also shown in (c) is the function $(-\partial f/\partial E)\times\Sigma\left(E\right)$ in Eq.~(\ref{eq:TDF}), at several temperatures in arbitrary units.
} \label{fig:Superdefect_effect}
\end{figure*}

\subsection{Relaxation times and their convergence}
\vspace{-10pt}
%
%
We first analyze the $e$-d RTs for vacancy defects in silicon, and study their convergence. 
Figure~\ref{fig:first_time_edRT}(a) gives the converged RTs (and their inverse, the scattering rates) of electrons and holes due to scattering with vacancy defects with a 1 ppm concentration.
As the electron and hole energy increases away from the respective band edges, the scattering rates increase (and the RTs decrease) 
due to the increased phase space for scattering $-$ that is, the larger number of final states that can be accessed in the scattering process. 
The latter is quantified by the density of states (DOS), which indeed shows a trend similar to the scattering rates. 
Overall, the $e$-d RTs are in the ps $-$ ns range, and thus much longer than the typical $e$-ph RTs near room temperature, which are in the fs $-$ ps range~\cite{Bernardi2016}. 
This result is consistent with the fact that $e$-ph scattering dominates near room temperature, while $e$-d scattering becomes important at low temperatures, where phonons are frozen out.\\
\indent
%
%
We find that the RTs depend on carrier energy, type of carrier, and type of neutral defect, at odds with simplified empirical models employed for decades to model $e$-d scattering~\cite{Erginsoy1950}. 
Figure~\ref{fig:first_time_edRT}(b) shows the RTs as a function of carrier energy in silicon for vacancy and interstitial defects, both with a 1 ppm concentration.
For the vacancy defects, the electron RT is roughly 1 ns at the conduction band minimum and the hole RT is roughly 300 ps at the valence band maximum; both these RTs decrease by an order of magnitude $\sim$100 meV away from the band edges. 
Overall, the holes relax significantly faster than the electrons, both near the band edge and at higher energies, a result we attribute to the higher density of states near the three-fold degenerate valence band maximum~\cite{Jhalani2017}. 
A similar asymmetry in the electron and hole RTs has been predicted recently in $e$-ph scattering in GaN~\cite{Jhalani2017}, where it was also attributed to the valence band degeneracy.\\
\indent 
For interstitials, in contrast, we find that the RTs are similar for electrons and holes. They are 150 ps for electrons and 100 ps for holes at the respective band edges, and for both carriers the RTs approach a value of 30 ps roughly 100 meV away from the band edges. 
The unexpected energy, carrier type, and defect type dependence of the RTs cannot be explained by the widely used Erginsoy formula~\cite{Erginsoy1950}, 
which predicts an energy-independent RT for $e$-d scattering due to neutral defects. 
Different from the Erginsoy model, our \textit{ab initio} calculations take the atomic and electronic structure into account, providing accurate results that are material and defect specific.\\
\indent
In \textit{ab initio} $e$-d calculations, there are significant challenges with converging the RTs, which have so far not been examined in detail. 
This convergence is crucial since many transport properties and physical observables associated with $e$-d interactions depend sensitively on the RTs. 
Figure~\ref{fig:Superdefect_effect} shows how to systematically converge the $e$-d RTs with respect to three key factors $-$ the supercell size, structural relaxation, 
and BZ grid used in the sum over final states in Eq.~(\ref{eq:incoherent}). This convergence study is discussed here for the vacancy case, although we find similar results for interstitial defects.\\
\indent
Figure~\ref{fig:Superdefect_effect}(a) shows the convergence of the RTs with respect to supercell size. Results are given for supercell sizes ranging from $4\,\times\,4 \,\times\, 4$ to $8\,\times\,8 \,\times\, 8$, in each case containing one vacancy at the center of the supercell. To isolate the role of supercell size, the atomic structure is not relaxed in these calculations.
The RTs in the $4\,\times\,4 \,\times\, 4$ and $6\,\times\,6 \,\times\, 6$ supercells are within $20$\% and $5$\%, respectively, of the $8\,\times\,8 \,\times\, 8$ supercell results, which can be considered fully converged. 
To verify that the RTs do not depend on defect position, we compute the RTs for an off-center vacancy that is placed away from the center of a $6\,\times\,6 \,\times\, 6$ supercell.  
For the same supercell size, the RTs of the off-center and centered vacancy match exactly, as they should.\\
\indent
Structural relaxation can be extensive around a defect and is expected to play an important role in accurately computing $e$-d interactions. 
Since it is costly to relax the structure in large supercells, an approximate scheme that retains accurate RTs is desirable. 
To this end, we compute the RTs in a supercell of a fixed size (here, $6\,\times\,6 \,\times\, 6$) in which only the atoms up to \textit{i}-th nearest neighbor shell of the vacancy defect are relaxed, and those that are farther away are kept fixed. 
Figure~\ref{fig:Superdefect_effect}(b) shows the RTs for structural relaxation up to the 1st, 2nd and 3rd nearest neighbor shell. 
We find that the RTs are almost converged for structural relaxation within the 2nd shell, and nearly identical to those for relaxation up to the 3rd shell, 
which can be considered converged since the atomic forces are negligible outside the 3rd nearest neighbor shell. 
The conclusion is that one needs to relax only a small portion of the atoms around the defect to accurately compute the $e$-d RTs.\\
\indent
Most critical when computing the $e$-d RTs is converging the $\bkp$-point grid in the sum over final states in Eq.~(\ref{eq:incoherent}), 
which is equivalent to converging the grid of transferred momenta, $\bq=\bkp-\bk+\bG$. 
There is a cross-convergence effect between this grid and the energy broadening $\eta$ employed to represent the delta function in Eq.~(\ref{eq:incoherent}), which is implemented as a normalized Gaussian with broadening, $\delta_\eta (x)= \frac{1}{\sqrt{2\pi}\eta}e^{-x^2/2\eta^2}$. 
The situation is fully analogous to converging the $e$-ph scattering rates~\cite{Zhou2016}. Briefly, the broadening has to be small enough to not alter the final result, but the smaller the broadening the denser the $\bkp$-point BZ grid needed to converge the sum in Eq.~(\ref{eq:incoherent}). Systematic convergence is achieved by starting with a small broadening (say, $\eta\approx10$ meV) and converging the $\bkp$-point BZ grid, and then decreasing the broadening to a smaller value and converging the BZ sum again. At convergence, the RTs do not change upon decreasing the broadening and converging the BZ sum. 
Note that the $\bkp$-point grid can in principle be distinct from the $\bk$-point grid at which the RTs are computed, but this is feasible in practice only if one has a mechanism to effectively interpolate the matrix elements. When this is possible, using random or importance sampling $\bkp$-point grids can significantly speed up the calculations~\cite{Bernardi2016}. Here, in each calculation, we use the same uniform BZ grid for $\bk$- and $\bkp$-points, and refer to it below as the BZ grid (a uniform $M\times M \times M$ grid will be denoted as an $M^{3}$ grid).\\
\indent
Figure~\ref{fig:Superdefect_effect}(c) shows the RTs for several values of the energy broadening $\eta$  
and gives the corresponding BZ grid at convergence.  
The BZ grid required to converge the RTs are denser for smaller values of the broadening; for $\eta$ values of 1, 5 and 15 meV, uniform BZ grids with $400^{3}$, $200^{3}$ and $150^{3}$ points are needed, respectively. 
For electron energies higher than $25$ meV above the conduction band minimum, a $15$ meV broadening and a $150^{3}$ BZ grid are sufficient to converge the RTs.  
For electron energies within 25 meV of the band edge, a 5 meV broadening with a $200^{3}$ BZ grid gives the same RTs 
as a smaller 1 meV broadening with a $400^{3}$ BZ grid~\footnote{Converging the RTs within a few meV of the band edges may require even smaller broadenings and denser BZ grids, but it is not necessary for mobility calculations, even at very low temperatures, because the carrier velocity vanishes at the band edges.}. 
The broadening and BZ grid values at convergence are similar for electrons and holes, and for vacancies and interstitials.\\
\indent
Importantly, the RTs computed with unconverged grids can differ widely from the converged values, 
especially at energies near the band edges [see Fig.~\ref{fig:Superdefect_effect}(c)], which critically contribute to charge transport. 
It is therefore essential to have an efficient method for computing and converging the $e$-d RTs on fine BZ grids to accurately compute charge transport at low temperature. 
%

%
\subsection{Defect-limited carrier mobility}
\vspace{-10pt}
At room temperature, where $e$-ph interactions typically dominate, charge transport 
can be accurately predicted from first principles in several families of materials~\cite{Mustafa2016, Park2014, Zhou2016, Lee2018, Zhou2018}.
However, many devices and experiments operate at low temperature, where charge transport is governed by $e$-d interactions. 
It is thus critically important to develop \textit{ab initio} calculations that can predict carrier dynamics at low temperature in the presence of $e$-d scattering. 
We compute the defect-limited carrier mobility $\mu$ at temperature $T$ within the RT approximation of the Boltzmann transport equation~\cite{Bernardi2016}: 
\begin{equation}\label{eq:Boltz}
\mu_{\alpha\beta} (T) =\frac{e}{n_{c}}\int_{-\infty}^{+\infty}\!\!dE\, \left[\, -\partial f(T,E)/\partial E \,\right] \times \Sigma_{\alpha\beta}\left(E\right),
\end{equation}
where $e$ is the electron charge, $n_{c}$ the carrier concentration, $f(T,E)$ the Fermi-Dirac distribution, and $\Sigma\left(E\right)$ the transport distribution function (TDF) at energy $E$, 
\begin{equation}\label{eq:TDF}
\Sigma_{\alpha\beta}\left(E\right)=\frac{2}{\Omega_{\rm uc}}\sum_{n\bk}\tau_{n\bk}\mathbf{v}_{n\bk}^{\alpha}\mathbf{v}_{n\bk}^{\beta}\delta\left(E-\varepsilon_{n\bk}\right),
\end{equation}
where $\alpha$ and $\beta$ are Cartesian directions. 
The TDF is computed with a tetrahedron integration method~\cite{Zhou2016}, using converged $e$-d RTs and Wannier-interpolated band velocities $\mathbf{v}_{n\bk}$~\cite{Yates2007,Mostofi2014}.
To estimate the carrier energy range contributing significantly to the mobility, we plot the integrand of the mobility formula in Eq.~(\ref{eq:Boltz}), the function $(-\partial f/\partial E)\times\Sigma\left(E\right)$, in Fig.~\ref{fig:Superdefect_effect}(c) for temperatures of $2$, $10$ and $160$ K.
As the temperature increases, the peak of the function broadens and moves away in energy from the band edges, 
indicating that the energy region contributing to the mobility shifts to higher carrier energies. 
The most stringent conditions for computing the RTs are below $10$ K, where the contribution to the mobility peaks $5$ meV away from the band edge; in this regime, BZ grids as dense as $200^{3}$ $\bkp$-points and a broadening of 5 meV are needed to accurately compute the mobility.\\
%
%
\begin{figure}[t!]
\centering
\includegraphics[width=1.0\linewidth]{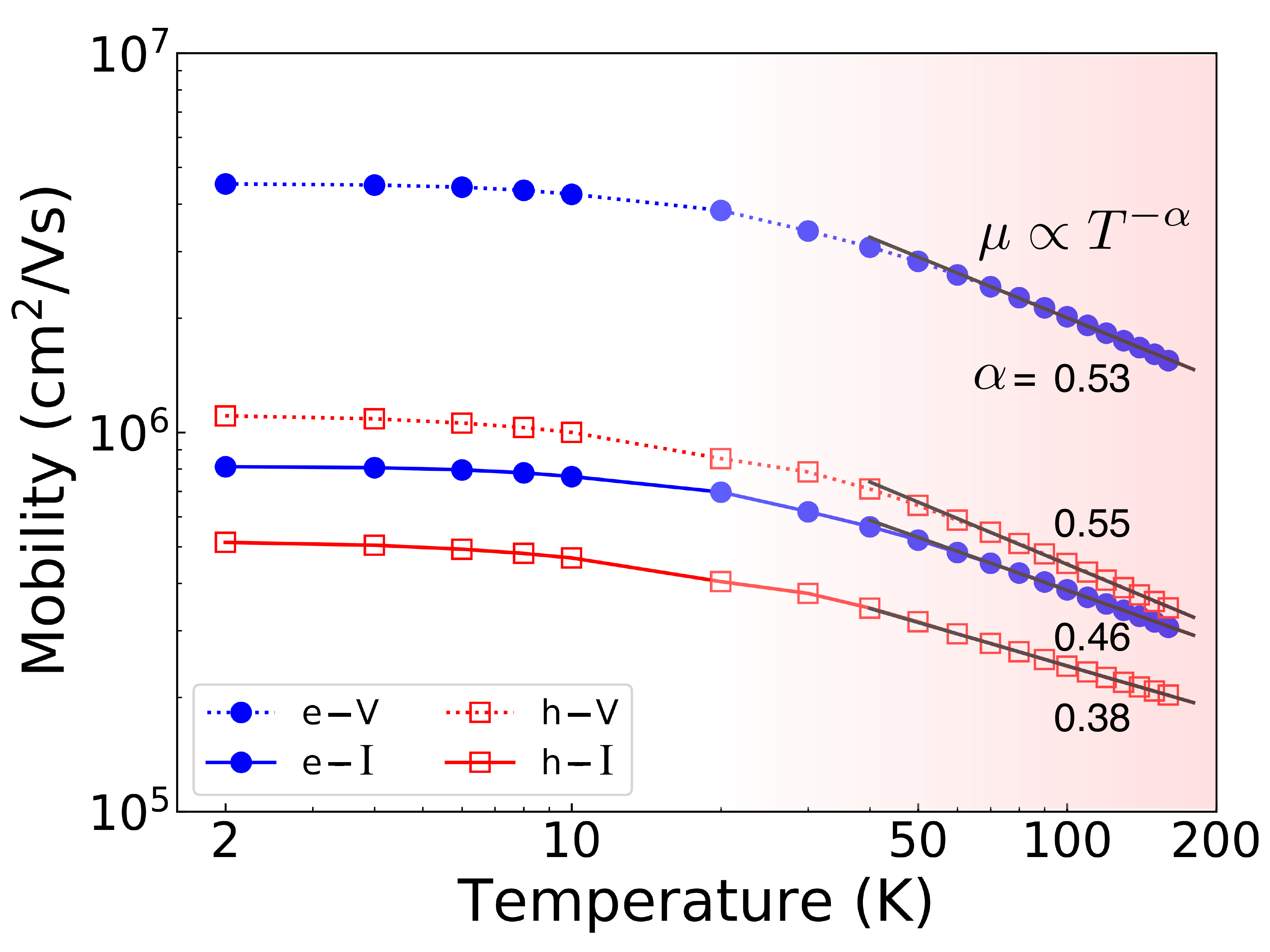}
\caption{Defect-limited mobilities in silicon, as a function of temperature below $150$ K. Shown are the results for electron-vacancy (e$-$V), hole-vacancy (h$-$V), electron-interstitial (e$-$I), and hole-interstitial (h$-$I) interactions.
The electron (solid circles) and hole (empty squares) mobilities are given for vacancy (dotted line) and interstitial (solid line) defects. 
The temperature dependence of the mobility above 50 K follows approximately a power law, $\mu \propto T^{-\alpha}$, with coefficients $\alpha$ given in figure. 
A defect concentration of 1 defect in 10$^6$ atoms is assumed. 
}\label{fig:Mobility}
\end{figure}
\indent
We compute the mobility of electrons and holes in silicon, considering separately vacancy and interstitial defects. 
Figure~\ref{fig:Mobility} shows the computed mobility curves at temperatures below 150 K. 
The electron and hole mobilities for vacancy defects are higher than the corresponding mobilities for interstitial defects due to the longer $e$-d RTs for vacancies. 
We find that, in all cases, the defect-limited mobility is roughly constant below 10 K, and decreases at higher temperatures. Note that $e$-ph interactions are not included here, so these trends are due solely to the $e$-d interactions. 
Above 50 K, the temperature dependence of the mobility is well approximated by a power law, $\mu \propto T^{-\alpha}$, with coefficients $\alpha$ of order 0.5.
For vacancy defects, the best-fit values of $\alpha$ are 0.53 for electrons and 0.55 for holes, and for interstitials 0.46 for electrons and 0.38 for holes.\\
\indent
An interesting interpretation, which is particularly apt for vacancies, is that the defects can be regarded as a substance added to the pure crystal to make an \lq\lq alloy''. 
One thus expects that the temperature dependence of the carrier mobility for defect scattering is similar to that of alloy scattering, for which a power law with $\alpha=0.5$ is expected based on existing models~\cite{Vaughan2012,hamaguchi2001basic}. Our results above 50 K are consistent with this interpretation, but we additionally find that the value of the exponent $\alpha$ depends on carrier and defect type. By contrast, the Erginsoy formula~\cite{Erginsoy1950} predicts a temperature independent mobility for neutral defect scattering, which is clearly inconsistent with our results, and also with experiment.\\ 
\indent
An early experiment~\cite{Norton1973} on $n$-type doped silicon obtained the mobility due to neutral impurity scattering by subtracting the lattice and ionized impurity contributions. 
The resulting mobility, which is limited by neutral defect scattering alone, decreases above 50 K, in agreement with our results. 
While this trend has been attributed by the authors to inelastic $e$-d scattering~\cite{Norton1973}, 
our results show that it can be explained by accurately computing elastic $e$-d scattering. 
The mobility decrease is due to the energy dependence of the RTs $-$ as the temperature increases, 
so does the average energy of the electrons contributing to the mobility [see Fig.~\ref{fig:Superdefect_effect}(c)], 
and their average RTs decrease as we have shown, causing the mobility decrease with temperature.\\ 
\indent
The mobility computed here using $e$-d interactions with only neutral point defects can be seen as an upper bound corresponding to an ideally pure material. 
In silicon, the mobility in the purest crystals (with impurity concentrations of $\sim$10$^{12}$ cm$^{-3}$) 
is roughly 10$^6$ cm$^2$/Vs at 10 K, a value that can be explained by ionized impurity scattering~\cite{Canali1975}. 
In these pure silicon samples, the concentration of neutral defects can also be as low as 10$^{12}$ cm$^{-3}$ ($C_{\rm d} \approx 10^{-10}$ in dimensionless units)~\cite{Canali1975}. 
We extend our results to this lower neutral defect concentration, and estimate a mobility limit in silicon of $\sim$10$^{10}$ cm$^2$/Vs at 10 K for  
an ideal scenario in which ionized impurity scattering is absent and only neutral point defects scatter the carriers. 
This mobility limit is higher than the value measured in samples with ionized impurity scattering, 
since the latter is much stronger than neutral defect scattering due to its long-range character. 
More extensive quantitative comparisons between computed and measured low-temperature charge transport data will be the subject of future work.
We conclude that our approach is a powerful tool to compute charge transport at low temperature and estimate mobility limit values in materials.

\section{Discussion}\label{sec:discussion}
\vspace{-10pt}
We close by discussing technical remarks and future research directions. 
Several improvements are possible to our calculations of $e$-d RTs. Similar to $e$-ph calculations, interpolating the $e$-d matrix elements, for example using Wannier functions or atomic orbitals~\cite{Agapito2018}, would be desirable, since one could compute the primitive cell wavefunctions only on coarse grids rather than on the fine grids needed to converge the RTs. 
In addition, since the broadening needed to converge the RTs increases with increasing carrier energy, 
using an adaptive broadening scheme could significantly speed up the RT calculations. 
One could use relatively coarse BZ grids and larger broadening values at higher carrier energies to save computational time, while using finer grids (and a smaller broadening) only at low carrier energy.\\
\indent
The method presented in this work can be extended in several ways. 
While our calculations focus on neutral point defects, our method can be generalized to treat charged defects, 
by adding their long-range Coulomb interaction to the local perturbation potential in reciprocal space, similar to what is done in $e$-ph calculations~\cite{Zhou2016}. 
The method can also be extended, using spin-polarized DFT calculations, to treat $e$-d scattering processes involving spin and spin-orbit coupling. 
Example applications include spin-flip processes due to magnetic impurities and defects in topological materials.\\
\indent
The proposed $e$-d calculations are general since they take into account the atomic structure of the material (including important structural relaxation effects around the defect) and its electronic structure. 
Unlike empirical models, there is no particular extension needed to treat different types of point and extended defects or different materials, provided one can afford the large DFT calculations needed to obtain the perturbation potentials. For example, our method is suitable for extended defects such as dislocations or grain boundaries, but to study them one may need supercells with thousands of atoms.\\ 
\indent
While this work focuses on the lowest order of perturbation theory, by including higher order $e$-d interactions one could investigate a wide range of low-temperature phenomena, including weak localization or antilocalization and universal conductance fluctuations~\cite{bruus2004many,mahan2013many}.
Our method can also be a starting point for efficient inelastic $e$-d scattering calculations~\cite{Barmparis2015}. 
\textit{Ab initio} calculations of $e$-d interactions are still in their infancy, and more work is needed to develop their potential and expand their scope.

\section{Conclusion} \label{sec:conclusion}
\vspace{-10pt}
We have presented an efficient approach that overcomes main technical challenges for \textit{ab initio} calculations of $e$-d interactions. 
The method is applied to compute and systematically converge the elastic $e$-d RTs 
and the associated defect-limited carrier mobility below $150$ K for vacancy and interstitial defects in silicon. 
The RTs exhibit a pronounced dependence on energy, carrier and defect type, and the defect-limited mobility is temperature dependent. 
These results cannot be explained using widely used empirical models of $e$-d interactions.
Our approach can provide new microscopic insight into $e$-d scattering processes. 
It is broadly applicable and can be generalized to treat charged defects, magnetic impurities, and extended defects.
We expect that this work will lay a solid foundation for efficient \textit{ab initio} calculations of $e$-d interactions. 
\vspace{-10pt}
%
\begin{acknowledgments}
\vspace{-10pt}
This work was supported by the Air Force Office of Scientific Research through the Young Investigator Program Grant FA9550-18-1-0280. 
J.-J. Z. was supported by the National Science Foundation under Grant No. ACI- 1642443, which provided for code development. 
This research used resources of the National Energy Research Scientific Computing Center, a DOE Office of Science User Facility supported by the Office of Science of the U.S. Department of Energy under Contract No. DE-AC02-05CH11231.
I-T. L. thanks Dr.~Luis Agapito, Dr.~Davide Sangalli, Vatsal Jhalani, Jinsoo Park and Xiao Tong for fruitful discussions.
\end{acknowledgments}

\appendix

\section{\uppercase{elastic and incoherent electron-defect scattering rate}}\label{sec:RTderivation}
\vspace{-10pt}
The scattering rate $\Gamma_{n\bk}$ can be written using Fermi's golden rule within lowest-order perturbation theory:
\begin{equation}\label{eq:deriveIncoS}
\Gamma_{n\bk} = \frac{2\pi}{\hbar}\sum_{n'\bkp}\left|\mel{\psi_{n'\bkp}}{\Delta \hat{H}}{\psi_{n\bk}}\right|^{2}\delta\left(\varepsilon_{n'\bkp}-\varepsilon_{n\bk}\right),
\end{equation}
where the perturbation $\Delta\hat{H}$ is the difference between the Kohn-Sham Hamiltonian $\hat{H}^{(\rm d)}$ of a crystal containing $N_{\rm d}$ identical defects and the Hamiltonian $\hat{H}^{(\rm p)}$ of the same crystal with no defects, namely, $\Delta\hat{H} = \hat{H}^{(\rm d)}-\hat{H}^{(\rm p)}$. 
The crystal is made up by $N_{\bk}$ primitive cells, and we apply Born-von Karman (BvK) periodic boundary conditions; the crystal volume is $\Omega_{\rm BvK} = N_{\bk} \Omega_{\rm uc}$, where $\Omega_{\rm uc}$ is the volume of the primitive cell. Above, $\ket{\psi_{n\bk}}$ are unperturbed Bloch wavefunctions with energy $\varepsilon_{n\bk}$, which in coordinate space read:
\begin{equation}\label{eq:psiform}
\braket{\br | \psi_{n\bk}}=\frac{1}{\sqrt{N_{\bk}}}\braket{\br | n\bk} =\frac{1}{\sqrt{N_{\bk}}}u_{n\bk}\left(\br\right)e^{i\bk\cdot\br},
\end{equation}
where $\ket{n\bk}$ is the Bloch wavefunction without the prefactor, and $u_{n\bk}\left(\br\right)$ is the periodic part of the Bloch wavefunction, normalized in the primitive cell as
\begin{equation}
\int_{\Omega_{\rm uc}}d^{3}r\ u^{*}_{n\bk}\left(\br\right) u_{n\bk}\left(\br\right) = 1.
\end{equation}
Since the kinetic energy is the same in the pristine and defect-containing systems, the difference of their Hamiltonians equals the sum of the perturbations due to all defects:
\begin{equation}\label{eq:potimp}
\Delta\hat{H}=\hat{H}^{(\rm d)}-\hat{H}^{(\rm p)} = \sum_{i=1}^{N_{\rm d}}\Delta V_{\rm e-d} (\br - \br_i),
\end{equation} 
where $\Delta V_{\rm e-d}(\br - \br_i)$ denotes the perturbation potential due to a defect located at $\br_i$, and we consider non-interacting defects of the same kind. Assuming that the scattering events are independent, we can write the scattering rate for elastic and incoherent scattering processes due to all defects as
\begin{equation}\label{eq:append_gamma}
\Gamma_{n\bk}=\frac{2\pi}{\hbar}\frac{n_{\rm at}C_{\rm d}}{N_{\bkp}}\sum_{n'\bkp}\left|M_{n'\bkp,n\bk}\right|^{2}\delta\left(\varepsilon_{n'\bkp}-\varepsilon_{n\bk}\right),
\end{equation}
where $n_{\rm at}$ is the number of atoms in the primitive cell,
$C_{\rm d}$, which is formally equal to $N_{\rm d}/\left(N_{\bk}\times n_{\rm at}\right)$, is in practice an assumed value of the defect concentration, and
$M_{n'\bkp,n\bk}$ is defined as the $e$-d matrix element for the perturbation due to a single defect:
\begin{equation}
M_{n'\bkp,n\bk}=\mel{n'\bkp}{\Delta V_{\rm e-d}}{n\bk}.
\end{equation}
Within our approximations, the scattering rate is proportional to the defect concentration, and can be computed at any desired value of the defect concentration, 
provided that the scattering events remain uncorrelated and the defects non-interacting throughout the concentration range of interest. 

\section{\uppercase{local matrix elements}}\label{sec:LocMderivation}
\vspace{-10pt}
The local matrix elements can be written as:
\begin{equation}\label{eq:deriveloc}
\begin{split}
&M_{n'\bkp,n\bk}^{\rm L}= \mel{n'\bkp}{\Delta V_{\rm L}\left(\br\right)}{{n\bk}}\\ 
&=\int_{\Omega_{\text{BvK}}}d^{3}r\ u_{n'\bkp}^{*}\left(\br\right)e^{-i\bkp\cdot\br}\Delta V_{\rm L}\left(\br\right)u_{n\bk}\left(\br\right)e^{i\bk\cdot\br}.
\end{split}
\end{equation}
We define the forward and inverse Fourier transforms of the local perturbation potential, respectively, as
\begin{equation}\label{eq:FTlocal}
\Delta V_{\rm L}\left(\br\right)=\frac{1}{N_{\bk}}\sum_{\bq}\Delta \widetilde{V}_{\rm L}\left(\bq\right)e^{i\bq\cdot\br}
\end{equation}
and
\begin{equation}\label{eq:InvFTlocal}
\begin{split}
\Delta \widetilde{V}_{\rm L}\left(\bq\right)=&\frac{1}{\Omega_{\rm uc}}\int_{\Omega_{\rm BvK}}d^{3}r\ \Delta V_{\rm L}\left(\br\right)e^{-i\bq\cdot\br}\\
                                            \approx& \frac{1}{\Omega_{\rm uc}}\int_{\Omega_{\rm sup}}d^{3}r\ \Delta V_{\rm L}\left(\br\right)e^{-i\bq\cdot\br},
\end{split}
\end{equation}
where in the last line we replace the crystal volume $\Omega_{\rm BvK}$ with the supercell volume $\Omega_{\rm sup}$, using the fact that the local perturbation potential vanishes at the supercell boundary,  
which is typically the case for supercells larger than a few primitive cells due to the localized nature of the perturbation potential. 
Inserting Eq.~(\ref{eq:FTlocal}) into Eq.~(\ref{eq:deriveloc}) and using the translational invariance of $u_{n\bk}\left(\br\right)$, we have
\begin{equation}\label{eq:localMmid}
M_{n'\bkp,n\bk}^{\rm L}=\sum_{\bG}\Delta \widetilde{V}_{\rm L}(\bkp-\bk+\bG)\mel{u_{n'\bkp}}{e^{i\bG\cdot\br}}{u_{n\bk}}_{\rm uc},
\end{equation}
with the plane-wave matrix elements defined as
\begin{equation}
\mel{u_{n'\bkp}}{e^{i\bG\cdot\br}}{u_{n\bk}}_{\rm uc} = \int_{\Omega_{\rm uc}}d^{3}r\ u^{*}_{n'\bkp}\left(\br\right) e^{i\bG\cdot\br} u_{n\bk}\left(\br\right),
\end{equation}
where $\bG$ are reciprocal lattice vectors of the primitive cell.
This formula is valid for any basis set. Here we use a plane wave basis for $u_{n\bk}\left(\br\right)$, and write
\begin{equation}
u_{n\bk}\left(\br\right)=\frac{1}{\sqrt{\Omega_{\rm uc}}}\sum_{\bG}C_{n\bk}\left(\bG\right)e^{i\bG\cdot\br},
\end{equation}
where $C_{n\bk}\left(\bG\right)$ is the Fourier coefficient of $u_{n\bk}\left(\br\right)$ at the reciprocal lattice vector $\bG$. 
The local matrix element formula in a plane wave basis becomes:
\begin{equation}
\begin{split}
&M_{n'\bkp,n\bk}^{\rm L}=\sum_{\bG}\Delta\widetilde{V}_{\rm L}(\bkp-\bk+\bG)\\
&\times \left[\sum_{\mathbf{G^{''}}}\sum_{\mathbf{G^{'}}}C_{n'\bkp}^{*}(\mathbf{G^{''}})\,C_{n\bk}(\mathbf{G^{'}})\,\delta_{\mathbf{G^{''}},\mathbf{G^{'}}+\bG}\right].
\end{split}
\end{equation}
This is the formula implemented in our code and used in this work.

\section{\uppercase{nonlocal matrix elements}}\label{sec:NonLocMderivation}
\vspace{-10pt}
To obtain an expression for the nonlocal matrix elements, we focus on the scalar product
\begin{equation}\label{eq:nonlocalbraket}
\begin{split}
&\braket{\beta_{j}^{(s)}\,|\,n\bk}=\int_{\Omega_{\rm BvK}}d^{3}r\beta_{j}^{*}\left(\br-\mathbf{\tau}_{s}\right)e^{i\bk\cdot\br}u_{n\bk}\left(\br\right)\\
&=\int_{\Omega_{\rm BvK}}d^{3}r\left[\frac{\beta_{j}\left(\br-\mathbf{\tau}_{s}\right)}{\sqrt{\Omega_{\rm uc}}}e^{-i\bk\cdot\br}\right]^{*}\sqrt{\Omega_{\rm uc}}u_{n\bk}\left(\br\right).
\end{split}
\end{equation}
We first fix the atomic position at the origin (by setting $\tau_{s}=0$) and then generalize the result to arbitrary atomic positions. 
We define generalized forward and inverse Fourier transforms of the Kleinman-Bylander (KB) projectors, respectively, as: 
\begin{equation}\label{eq:FTnonlocal}
\frac{\beta_{j}\left(\br\right)}{\sqrt{\Omega_{\rm uc}}}e^{-i\bk\cdot\br}=\frac{1}{\Omega_{\rm BvK}}\sum_{\bq}B_{j\bk}\left(\bq\right)e^{i\bq\cdot\br}
\end{equation}
and
\begin{equation}
\begin{split}
B_{j\bk}\left(\bq\right)&=\int_{\Omega_{\rm BvK}}d^{3}r\left[\frac{\beta_{j}\left(\br\right)}{\sqrt{\Omega_{\rm uc}}}e^{-i\bk\cdot\br}\right]e^{-i\bq\cdot\br}\\
&=\int_{\Omega_{\rm sup}}d^{3}r\left[\frac{\beta_{j}\left(\br\right)}{\sqrt{\Omega_{\rm uc}}}e^{-i\bk\cdot\br}\right]e^{-i\bq\cdot\br},
\end{split}
\end{equation}
where we replace the crystal volume $\Omega_{\rm BvK}$ with the supercell volume $\Omega_{\rm sup}$ because the KB projector is localized around the core atomic region. If a general atomic position $\mathbf{\tau}_{s}$ is chosen, the Fourier coefficient $B_{j\bk}^{(s)}\left(\bq\right)$ becomes, using the properties of the Fourier transforms,
\begin{equation}
B_{j\bk}^{(s)}\left(\bq\right)=e^{-i(\bk+\bq)\cdot\mathbf{\tau_{s}}}B_{j\bk}\left(\bq\right).
\end{equation}
The scalar product in Eq.~(\ref{eq:nonlocalbraket}), after inserting Eq.~(\ref{eq:FTnonlocal}) into Eq.~(\ref{eq:nonlocalbraket}), becomes
\begin{equation}
\braket{\beta_{j}^{(s)}\,|\,n\bk}
=\frac{1}{\sqrt{\Omega_{\rm uc}}}\sum_{\bG}B_{j\bk}^{(s)*}\left(\bG\right)\braket{e^{i\bG\cdot\br}\,|\,u_{n\bk}}_{\rm uc},
\end{equation}
where we used the translational invariance of $u_{n\bk}\left(\br\right)$. This formula is valid for any basis set. Here we use a plane wave basis, 
so the matrix elements of $\hat{V}_{\rm NL}$, for each of the pristine and defect-containing supercells (labelled by $\alpha$ = d, $\!$p), read: 
\begin{equation}
\begin{split}
&\braket{n'\bkp | \hat{V}_{\rm NL}^{\rm (\alpha)} | n\bk} = \sum_{s_\alpha=1}\sum_{ij} D_{ij}^{(s_\alpha)} \times \\ 
&\left[\sum_{\mathbf{G^{'}}}B_{i\bkp}^{(s_\alpha)}(\mathbf{G^{'}})\,C_{n'\bkp}^{*}(\mathbf{G^{'}})\right]
 \left[\sum_{\bG}B_{j\bk}^{(s_\alpha)*}(\bG)\,C_{n\bk}(\bG)\right],
\end{split}
\end{equation}
where we use the same notation as in Eq.~(\ref{eq:nonlocalM}). The nonlocal matrix elements are computed as the difference in Eq.~(\ref{eq:totnonlocal}). This is the formula implemented in our code and used in this work.\\
\bibliography{perturbo-ed}
\end{document}